\documentclass[aps,prd,showpacs,preprintnumbers,groupedaddress,12pt]{revtex4}
\usepackage{amsmath}
\usepackage{graphicx}
\usepackage{bm}
\usepackage{float}
\usepackage{color,colordvi}
\usepackage{feynarts}
\begin{document}
\def\cM{{\cal M}} 
\def\cO{{\cal O}}
\def\cK{{\cal K}}
\def\cS{{\cal S}}
\newcommand{\mh}{m_h}
\newcommand{\mw}{m_W}
\newcommand{\mz}{m_Z}
\newcommand{\mt}{m_t}
\newcommand{\mb}{m_b}
\def\ga{\mathrel{\raise.3ex\hbox{$>$\kern-.75em\lower1ex\hbox{$\sim$}}}}
\def\la{\mathrel{\raise.3ex\hbox{$<$\kern-.75em\lower1ex\hbox{$\sim$}}}}
\begin{flushright}
LPHEA--06--04\\
FSU-HEP-061010
\end{flushright}
\title{Searching for a {\sc{cp}}-odd Higgs via a pair \\ of gauge
    bosons at the {\sc{lhc}}}
\author{Abdesslam Arhrib$^1$} 
\email[]{aarhrib@ictp.it}
\author{Rachid Benbrik$^2$}
\email[]{r.benbrik@ucam.ac.ma}
\affiliation{$^1$D\'epartement de Math\'ematiques,
             Facult\'e des Sciences et Techniques,
             B.P. 416 Tanger, Morrocco.\\
             $^2$LPHEA, D\'epartement de Physique
             Facult\'e des Sciences-Semlalia
             B.P. 2390 Marrakech, Morocco.}
\author{Bryan Field}
\email[]{bfield@hep.fsu.edu}
\affiliation{Department of Physics, Florida State University,
       Tallahassee, Florida 32306-4350, USA}

\date{Sept 22, 2006}
\begin{abstract}

The \textsc{cp}-odd Higgs boson $A^0$ of the Minimal Supersymmetric
Standard Model (\textsc{mssm}) and Two Higgs Doublet Model
(2\textsc{hdm}) will usually decay into the heaviest possible fermion
-- antifermion pair available. The $A^0 \to VV$ decay where, $V
=\gamma, Z, W^\pm$, and gluons are of particular interest as they are
not allowed at tree level and hence they may offer information about 
the underlying new physics that enters at one loop level. 
In this paper all branching ratios of the
\textsc{cp}-odd Higgs boson $A^0$ both in the \textsc{mssm} and
2\textsc{hdm} are presented for this channel including all relevant
Standard Model (\textsc{sm}) and \textsc{mssm} particles. This
discovery channel might provide an opportunity to search for a
\textsc{cp}-odd Higgs boson at the Large Hadron Collider
(\textsc{lhc}) and new physics beyond the Standard Model. Expressions
for these decays are given in the Appendices.

\end{abstract}

\pacs{13.85.-t, 14.80.Bn, 14.80.Cp}
\maketitle
\newpage
\section{Introduction}
\label{sec:intro}

One of the main goals of future colliders such as the CERN
\textsc{lhc} and the future International Linear Collider
(\textsc{ilc}) is to study the Higgs sector of the Standard Model
(\textsc{sm})\cite{Weinberg:1967tq, Glashow:1961tr, Salam:1969zz}.
Moreover, the scalar sector of the \textsc{sm} can be enlarged with a
simple extension such as the Minimal Supersymmetric Standard Model
(\textsc{mssm}) and the Two Higgs Doublet Model
(2\textsc{hdm})\cite{Gunion:1989we, Djouadi:2005gi, Djouadi:2005gj},
which are studied in this paper. Both in the 2\textsc{hdm} and
\textsc{mssm} the electroweak symmetry breaking is generated by two
Higgs doublets fields $\Phi_{1,2}$, the imaginary parts of which
combine to produce a \textsc{cp}--odd Higgs boson $A^0$ and a
Goldstone mode $G^0$, both of which must have the same quantum
numbers. Consequently $A^0$ and $G^0$ have the quantum numbers
$0^{+-}$. Owing to the $J^{PC}(A^0)=0^{+-}$ the vertices $A^0 WW$ and
$A^0 ZZ$ are forbidden at tree level.

Another way to explain the absence of the $A^0 WW$ and $A^0 ZZ$
vertices is the following: these interactions come from the kinematic
term $(D_\mu\Phi)(D^\mu\Phi)^+$ after replacing one of the $\Phi$'s by
its \textsc{vev}, but in a \textsc{cp}--conserving model the
\textsc{vev} is real while $A^0$ comes from the imaginary part of
$\Phi^0$, therefore $A^0$ cannot couple to any massive vector
boson. Consequently these vertices can appear first at the one-loop
level. In this study, our concern is the \textsc{cp}-odd $A^0$ decay
into a pair of gauge bosons, and we will therefore review the
production mechanisms for $A^0$.

At future $e^+e^-$ machines, the \textsc{cp}-even Higgs $h^0$ is
mainly produced via the associated production with a $Z$ boson. At
high energies, $WW$ fusion to $h^0$ becomes significant. Due to its
\textsc{cp} nature, the \textsc{cp}--odd $A^0$ possesses no tree-level
coupling $A^0 ZZ$ and $A^0 WW$. Consequently, both of the above
processes, Higgsstrahlung and $WW$ fusion, are unavailable to the
$A^0$ at $e^+e^-$ machines. The dominant contribution is therefore
from higher order diagrams such as $e^+e^- \to \gamma
A^0$\cite{Djouadi:1996yq, Krawczyk:1998kk, Djouadi:1996ws}, $e^+e^-
\to Z A^0$\cite{Akeroyd:1999gu, Akeroyd:2001ak} and $e^+e^-\to A^0 \nu
\bar{\nu}$\cite{Arhrib:2002ti, Farris:2003pn}. The production rate of
those loop mediated process turns out to be quite small. Therefore,
the only realistic tree level mechanisms for \textsc{cp}-odd Higgs
bosons at $e^+e^-$ colliders \cite{Aguilar-Saavedra:2001rg,
Abe:2001gc, Abe:2001np} are via $e^+e^- \to h^0 A^0$ and $e^+e^- \to b
\bar{b} A^0, t \bar{t} A^0$ \cite{Djouadi:1991tk, Djouadi:1992gp,
Dawson:1998qq}. At $\gamma\gamma$ collider (muon collider), $A^0$ can
be copiously produced as a resonance $\gamma\gamma\to A^0$
($\mu^+\mu^- \to A^0$) as well as in association with a $Z$ boson
$\gamma \gamma \to Z A^0$ \cite{Gounaris:2001rk} ($\mu^+ \mu^- \to Z
A^0$\cite{Akeroyd:1999xf, Akeroyd:2000zs}).

The situation is different at hadron colliders. The main production
mechanism is through gluon-gluon fusion $g g \to A^0$ or $g g \to A^0
Q \bar{Q}$ (where $Q = t, b$)\cite{Gunion:1991cw}. Those processes can
provide an ample sample event at the \textsc{lhc}. Other mechanisms
like associate production of $A^0$ with a $Z$ boson is also possible
at the \textsc{lhc}\cite{Yin:2002sq}. From the experimental side,
current mass bounds from LEP--II for the neutral pseudoscalar $A^0$ of
the \textsc{mssm} is $m_{A^0} \ge 85$~GeV in the $m_{h}^{max}$
scenario\cite{Abbiendi:2004ww, Achard:2002zr}. Within 2\textsc{hdm},
the \textsc{opal} collaboration has used the pair-production $e^+e^-
\to h^0 A^0$ cross-section assuming 100\% decays into hadrons
independent of hadron flavor. It is found that the region $1 \lesssim
m_h \lesssim 55$ GeV and $3 \lesssim m_{A^0} \lesssim 63$~GeV is excluded
at 95\% CL independent of the choice of the 2\textsc{hdm}-II
parameters\cite{Abbiendi:2004gn}.

The aim of this paper is the study of the decay of the \textsc{cp}-odd
$A^0$ into a pair of gauge boson both in 2\textsc{hdm} and
\textsc{mssm}. Although these decays are rare processes, loop and/or
threshold effects could potentially give a substantial
effect. Moreover, once worked out, any experimental deviation from the
results within such a model should bring some fruitful information on
the new physics and allow one to distinguish between models. We would
like to mention also that these channels have a very clear signature
and might emerge easily at future colliders. For instance, if $A^0 \to
ZZ$ is enhanced enough, this decay may lead to the so-called
gold-platted event signature $A^0 \to ZZ \to l^+l^-l^+l^-$.

These decays have been already studied in literature both in the
\textsc{mssm}\cite{Gunion:1991cw, Chankowski:1992es} and
2\textsc{hdm}. The \textsc{mssm} contribution to $A^0 \to ZZ, W^+W^-$
has been first studied in \cite{Gunion:1991cw}, however only heavy
\textsc{sm} fermions were included. In \cite{Chankowski:1992es},
chargino and neutralino contributions to $A^0\to ZZ, W^+W^-$ were
included together with the \textsc{sm} fermions. However in
Ref.~\cite{Chankowski:1992es}, only $A^0 \to ZZ$ was considered
and the results was presented in terms of decay width only, 
no branching ratio was given.
At tree level, \textsc{cp}-odd pseudoscalar can decay, when its
kinematically possible, to a pair of standard model fermions
$f\bar{f}$, a pair of charginos, a pair of neutralinos, a pair of
scalar quarks (squarks), as well as to $Z^0 h^0$, $Z^0 H^0$ and $W^\pm
H^\mp$. On the other hand, decays modes like $A^0 \to gg$, $A^0 \to
\gamma \gamma$, $A^0 \to \gamma Z$, $A^0 \to ZZ$ and $A^0 \to W^+W^-$ are
mediated at one-loop level. The decays $A^0\to gg$, $A^0 \to \gamma
\gamma$, $A^0 \to \gamma Z$ has been evaluated by many groups and are
well implemented in \textsc{hdecay} program\cite{Djouadi:1997yw,
Djouadi:1995gv}. In this section, we would like to focus on the
evaluation of $A^0\to ZZ$ and $A^0\to W^+W^-$ both in 2\textsc{hdm} and
\textsc{mssm}.

We will update those analysis in the \textsc{mssm} by including both
\textsc{sm} fermions, charginos and neutralinos contributions and also
by presenting both the decay width and branching ratios. We will also
study those decays $A^0\to ZZ, W^+W^-$ in both type~I and type~II
2\textsc{hdm}.

In this paper, we give a complete calculation of the \textsc{cp}-odd
Higgs boson $A^0$ decay into $A^0 \to ZZ$, $A^0 \to W^+W^-$ in the general
\textsc{cp}-conserving 2\textsc{hdm} and the full \textsc{mssm}. In
Section~\ref{sec:cp-decay}, we present the relevant Feynman diagrams
and formulas involved in our calculation. In
Section~\ref{sec:numerics}, we present our numerical results and
discussions. Section~\ref{sec:conclusion} contains our conclusions.

\section{\textsc{cp}-odd decay: $A^0\to VV$}
\label{sec:cp-decay}

In the \textsc{mssm} or 2\textsc{hdm} with \textsc{cp} conservation,
the $A^0 ZZ$ and $A^0 WW$ couplings are forbidden at tree-level. They
can be generated at the one-loop level via the Feynman diagrams in
Fig.~\ref{fig:azz-diagrams} and Fig.~\ref{fig:aww-diagrams}. 
In Fig.~\ref{fig:azz-diagrams}, topologies like \ref{fig:azz-diagrams}.4
and \ref{fig:azz-diagrams}.5 exactly vanishes, 
since Z behaves like \textsc{cp}-odd there is no transition $Z$-$\phi$, $\phi=h^0$, $H^0$.
For the same reason, $A^0$ is \textsc{cp}-odd, there is no transition $A^0$-$\phi$ and then
topologies like \ref{fig:azz-diagrams}.6 vanishes.\\
In Fig.~\ref{fig:aww-diagrams}, topologies like \ref{fig:aww-diagrams}.8
and \ref{fig:aww-diagrams}.9 exactly vanishes for an on-shell $W$
boson since $W^\pm$-$H^\mp$ transition is proportional to the $W$
momentum. In topologies like \ref{fig:aww-diagrams}.11 
 the transition $A^0$-$V$ is proportional to
$A^0$ momentum, once contracted with $VW^+W^-$ coupling it vanishes
for an on-shell $W$ boson. Like in Fig.~\ref{fig:azz-diagrams}
there is no transition $A^0$-$\phi$ and then
topologies like \ref{fig:aww-diagrams}.10 vanishes.

In both cases, we did not include gauge bosons and scalars (scalar
bosons and scalar fermions) in the loops. As it is argued in
\cite{Gunion:1989we, Gunion:1991cw}, the sum of each of these
contributions must cancel. The reason is that \textsc{p} and
\textsc{c} are separately conserved in the bosonic sector before the
introduction of fermions.  Since the $A^0$ and the $Z$ boson are
\textsc{c}-odd and the \textsc{cp}-odd zero angular momentum state of
$W^+W^-$ must be \textsc{c}-even. Therefore, the coupling $A^0 VV$ is
forbidden to all order in the bosonic sector. When fermions are
introduced, \textsc{c}, \textsc{p} and \textsc{cp} are no longer
conserved and as a consequence $A^0 VV$ can be induced at higher loop
levels.

The above statement applies to the contribution of scalar fermions as
well. In this case, this can be easily be seen because \textsc{cp}-odd
$A^0$ couplings to scalar fermions satisfy the following relation $A^0
\widetilde{f}_i \widetilde{f}_j^* = - A^0 \widetilde{f}_i^*
\widetilde{f}_j$. Consequently, the total contribution of scalar
fermions to $A^0ZZ$ or $A^0W^+W^-$ exactly cancel when \textsc{cp} is
conserved.

Given the fact that $A^0$ is \textsc{cp}-odd, the effective Lagrangian for
$A^0VV$ must have the form:
\begin{equation}
{\cal L}_{A^0VV} = 
  g_{A^{0}VV} \epsilon^{\mu\nu\rho\delta} V_{\mu\nu} V_{\rho\delta}
\label{eqn:eff:lagrange}
\end{equation}
The effective coupling $g_{A^{0}VV}$ has a dimension -1, it is
expected to be of the form: $g_{A^{0}VV}=1/m_{W}\ F(M_{S})$, where $F$
is a dimensionless function of $M_{S}$ which behaves like $\log(M_S)$
and $M_{S}$ is the masses of internal particles in the
loops. Therefore, for $A^0 \to V_1V_2$ we expect only a logarithmic
dependence on the internal masses. The general one-loop amplitude
takes the following form
\begin{equation}
{\cal M}(A^0 \to V_1 V_2) =
\frac{g^3 N_C}{16 \pi^2 m_W}
\epsilon_{\mu\nu\rho\delta}
\epsilon_1^{\mu} \epsilon_2^{\nu} p_1^\rho p_2^\delta {\cal A}_{V_1 V_2},
\label{eqn:amp}
\end{equation}
where $N_C=3$ for quarks and $1$ for leptons and charginos neutralinos
in the internal loop. Analytical expression for ${\cal A}_{V_1V_2}$
both for \textsc{sm} fermions and charginos neutralinos loops is given
in Appendix~C. The partial decay is then computed from the above
amplitude and is given by:
\begin{equation}
\Gamma(A^0 \to V_1 V_2) = S_{12} \frac{g^6 N_C^2 \lambda^{3/2}}
{2^{13} \pi^5 m_W^2 m_{A^0}^3} |{\cal A}_{V_1V_2}|^2,
\label{eqn:gamma}
\end{equation}
where $S_{12}=1/2$ in case of identical particles in the final state
and $\lambda = (m_{A^0}^2 - m_{V_1}^2 - m_{V_2}^2)^2 - 4 m_{V_1}^2
m_{V_2}^2$.

Both in 2\textsc{hdm} in \textsc{mssm}, the total decay widths
$\Gamma_{A^0}^{ \rm{ 2 \textsc{hdm} } }$ and $\Gamma_{A^0}^{ \rm{
\textsc{mssm} } }$ are computed as follows:
\begin{equation}
\Gamma_{A^0}^{\rm{2\textsc{hdm}}} = 
\sum_{f}   \, \Gamma( A^0 \to f\bar{f}) +
\sum_{V_i} \, \Gamma( A^0 \to V_1V_2 ) + 
              \Gamma( A^0 \to Z \Phi^0 ) +
              \Gamma( A^0 \to W^\pm H^\mp ),
\label{eqn:width}
\end{equation}
where the summation in $\sum_{V_i} \Gamma( A^0\ \to V_1V_2)$ stand for $V_i
= \gamma$, g, $Z$ and $W$, $\Phi^0 = h^0$ or $H^0$. QCD corrections to
$A^0 \to f\bar{f}$ and $A^0 \to \{V^* \Phi\}$ decays are not included
in the widths. In the case of \textsc{mssm},
$\Gamma_{A^0}^{\rm{\textsc{mssm}}}$ is obtained from
$\Gamma_{A^0}^{\rm{2\textsc{hdm}}}$ by adding the decay of $A^0$ to
SUSY particles: $A^0 \to \tilde{\chi}_i^0 \tilde{\chi}_j^0$,
 $A^0 \to \tilde{\chi}_i^+ \tilde{\chi}_j^{-}$ and 
$A^0 \to \tilde{f}_i \tilde{f}_j^\star $.
\begin{figure}[t!]
\begin{center}
\input{azz1.tex}
\vspace{-10.2cm}
\caption{Generic contribution to $A^0 \to ZZ$ in 2\textsc{hdm} and
         \textsc{mssm}, F denotes any fermion particles, 
S denotes any scalar particles, P denotes anay particles 
in the model that can fit into the diagrams and 
$\phi$ is one of the \textsc{cp}-even scalars $h^0$ or $H^0$.}
\label{fig:azz-diagrams}
\end{center}
\end{figure}
\begin{figure}[t!]
\begin{center}
\input{aww1.tex}
\vspace{-7cm}
\caption{Generic contribution to $A^0 \to W^+W^-$ in 2\textsc{hdm} and
         \textsc{mssm}, F denotes any fermion particles, 
S denotes any scalar particles, P denotes anay particles 
in the model that can fit into the diagrams and 
$\phi$ is one of the \textsc{cp}-even scalars $h^0$ or $H^0$.}
\label{fig:aww-diagrams}
\end{center}
\end{figure}

There are two types of the 2\textsc{hdm}, depending on which Higgs
field is responsible for the masses of quarks and
leptons. Consequently, the Yukawa couplings of the quarks and leptons
to the Higgs bosons are different. For the $A^0$ we find
${\mathcal{L}} = -i g_{A^{0} ff}\bar{f} \gamma_5 f A^0$ where the
couplings constants are given in \textsc{table}~\ref{tabel1}.

\begin{table}[hbt!]
\begin{center}
 \renewcommand{\arraystretch}{1.30}
\begin{tabular}{||p{2.5cm}|p{2.5cm}||p{2.5cm}||}
\hline
\quad Couplings\quad &\quad\textsc{2HDM-I}\quad &\quad\textsc{2HDM-II}\quad\\
\hline
$\quad g_{A^0 t\bar{t}}\quad$& $\quad\frac{m_t}{v}\cot\beta\quad$ &
$\quad\frac{m_t}{v}\cot\beta\quad$\\
\hline
$\quad g_{A^0 b\bar{b}}\quad$& $\quad\frac{m_b}{v}\cot\beta\quad$ &
$\quad\frac{m_b}{v}\tan\beta\quad$\\
\hline
$\quad g_{A^0 \tau\bar{\tau}}\quad $& $\quad\frac{m_{\tau}}{v}\cot\beta\quad$ &
$\quad\frac{m_\tau}{v}\tan\beta\quad$\\
\hline
\end{tabular}
\end{center}
\caption{Couplings of \textsc{cp}-odd Higgs boson $A^0$ with fermions both in
  \textsc{2HDM-I}  and \textsc{2HDM-II}. The \textsc{mssm} 
is required to have 2\textsc{hdm}--II couplings.}
\label{tabel1}
\end{table} 
We have evaluated the one-loop induced process $A^0 \to V_1 V_2$ in the
`t~Hooft-Feynman gauge, and using dimensional
regularization\cite{'tHooft:1972fi, Breitenlohner:1977hr}. The types
of Feynman diagrams are depicted in Figure~\ref{fig:azz-diagrams} and
Figure~\ref{fig:aww-diagrams}. All the Feynman diagrams have been
generated and computed using FeynArts and FeynCalc\cite{Hahn:2000kx,
Hahn:2001rv, Hahn:1998yk, Kublbeck:1990xc} packages. We have also used
the fortran FF--package \cite{vanOldenborgh:1990yc, Hahn:1999mt} in
the numerical analysis.

\section{Numerics and Discussions}
\label{sec:numerics}

In our numerical evaluations, we use the following experimental input
quantities\cite{Eidelman:2004wy}: $\alpha^{-1} = 129$, $m_Z$, $m_W$,
$m_t$, $m_b=$ $91.1875$, $80.45$, $174.3$, $4.7$~GeV. In the
\textsc{mssm}, we specify the free parameters that will be used as
fellow: i) The \textsc{mssm} Higgs sector is parameterized by the
\textsc{cp}-odd mass $m_{A^0}$ and $\tan\beta$, taking into account
radiative corrections from \cite{Heinemeyer:1999be,
Heinemeyer:1998np}, and we assume $\tan\beta \gtrsim 3$. ii) The
chargino--neutralino sector can be parameterized by the gaugino-mass
terms $M_1$, $M_2$, and the Higgsino-mass term $\mu$. For
simplification $M_1 \approx M_2/2$ is assumed. iii) Sfermions are
characterized by a common soft-breaking sfermion mass $M_{SUSY} \equiv
\widetilde{M}_L = \widetilde{M}_R$, $\mu$ parameter and soft trilinear
couplings for third generation scalar fermions $A_{t,b,\tau}$.

When varying the \textsc{mssm} parameters, we take into account also
the following constraint The extra contributions to the $\delta\rho$
parameter from the Higgs scalars should not exceed the current limits
from precision measurements\cite{Eidelman:2004wy}: $|\delta\rho|
\lesssim 0.003$.
\begin{figure}
  \begin{tabular}{cc}
    \resizebox{88mm}{!}{\includegraphics{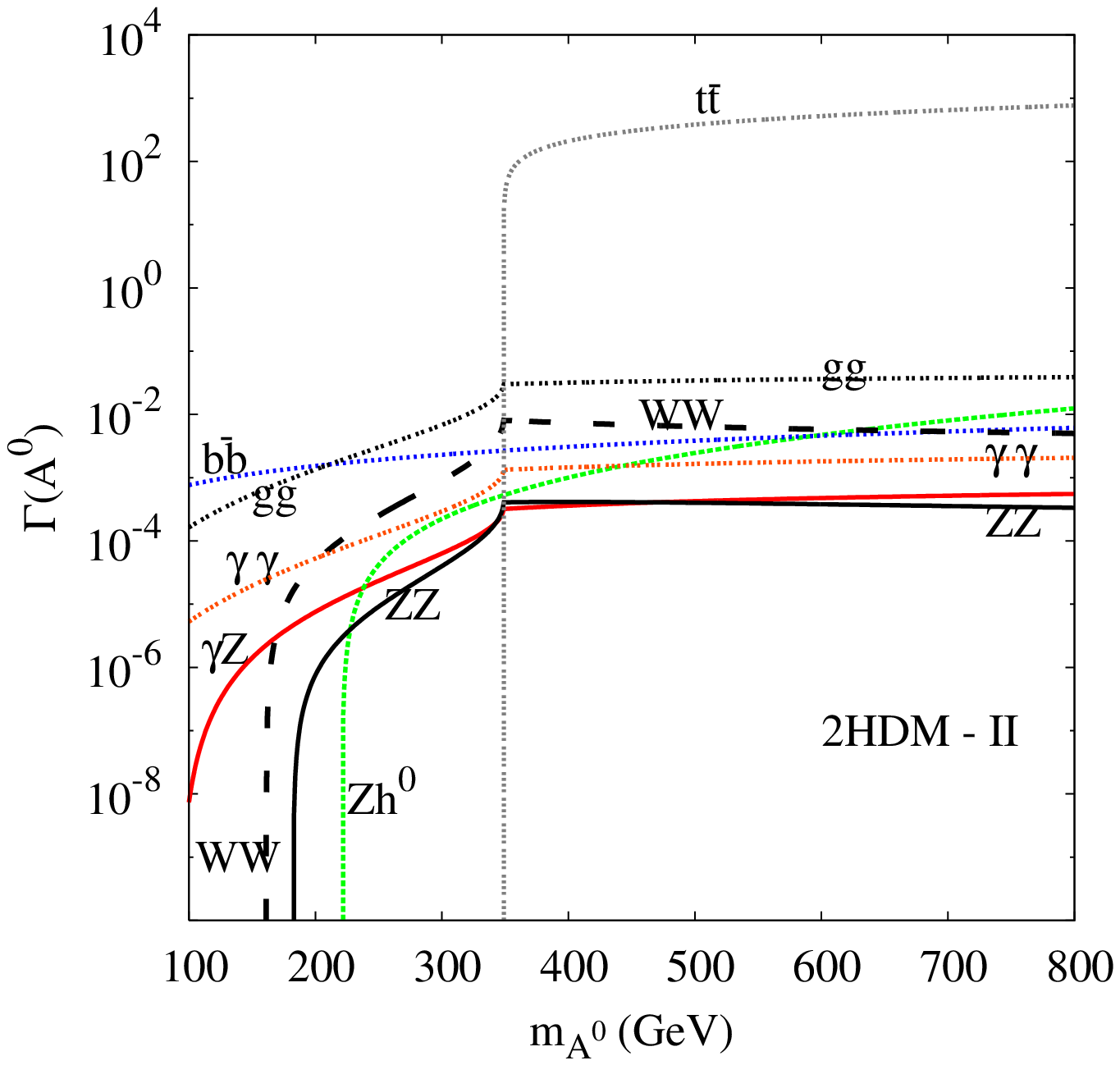}} &
    \resizebox{88mm}{!}{\includegraphics{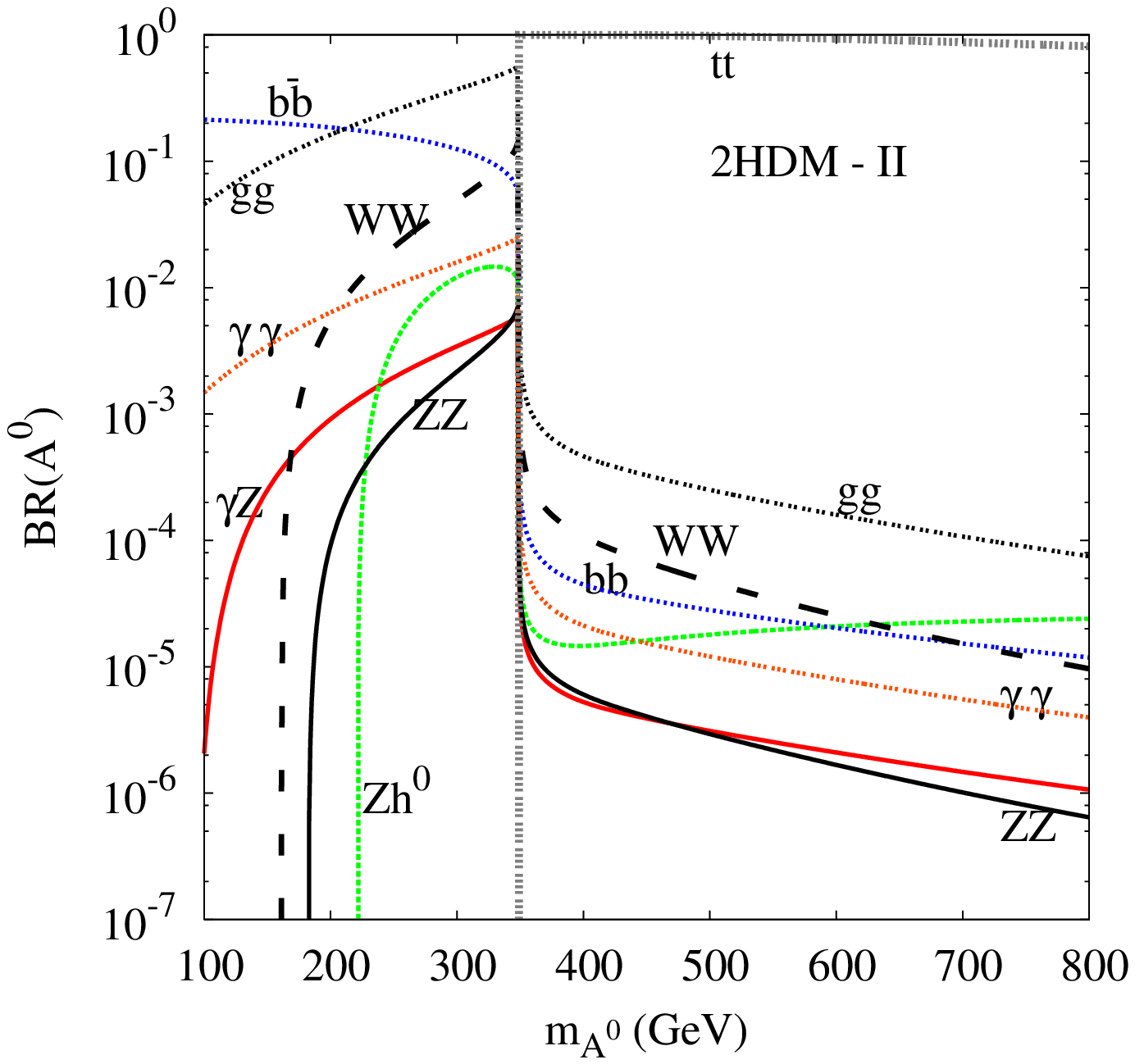}}
  \end{tabular}
\caption{Decays widths (left) and branching ratios (right) of 
 \textsc{cp}-odd $A^0$ as a
         function of $m_{A^0}$ in 2\textsc{hdm} Type II , 
 for the parameters $m_{h^0} = 130$ GeV,
         $m_{H^0}=345$ GeV, $m_{H^{\pm}}=340$ GeV, $\tan\beta = 0.42$.}
\label{fig:fig1}
\end{figure}
\begin{figure}
  \begin{tabular}{cc}
    \resizebox{88mm}{!}{\includegraphics{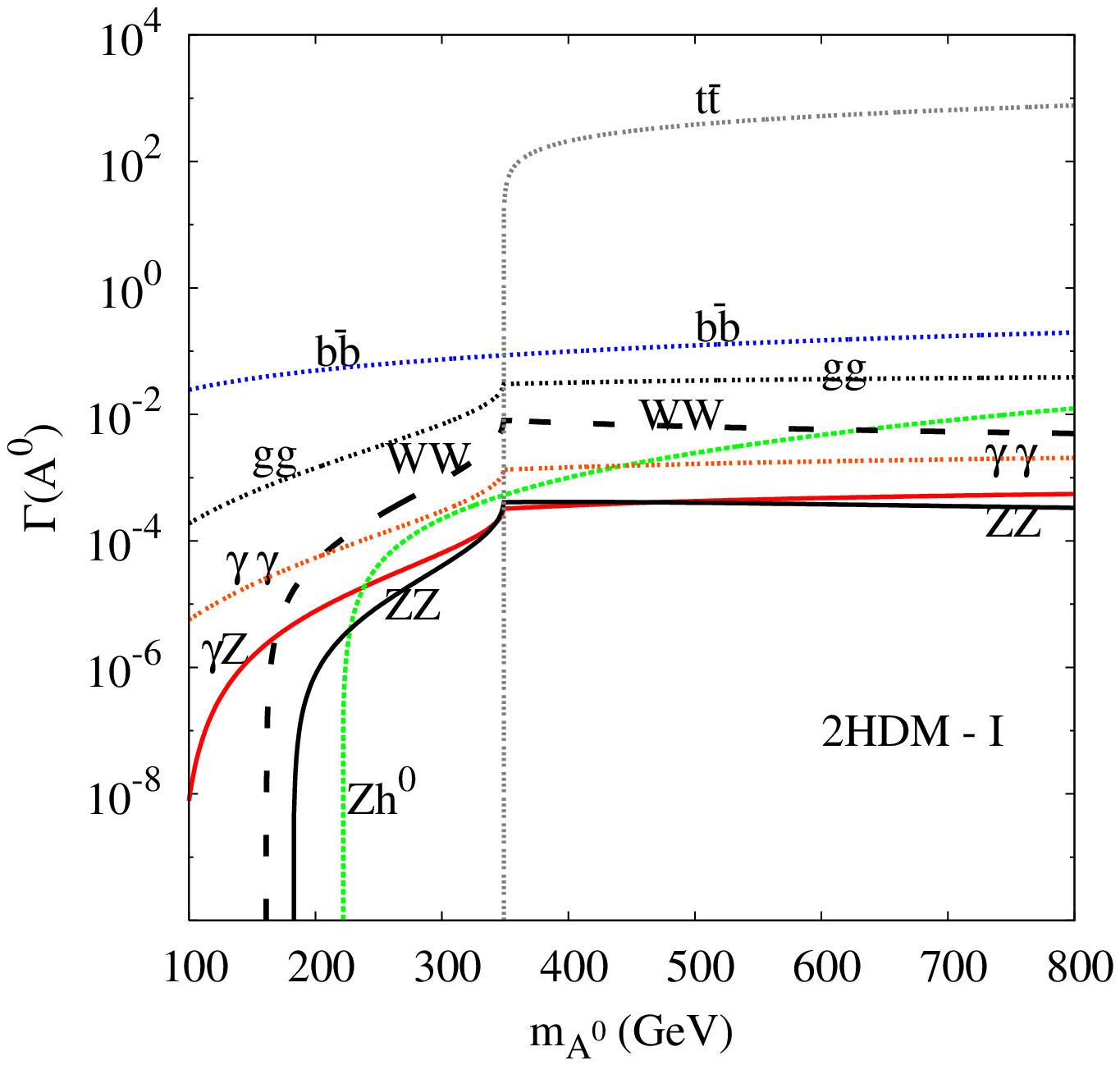}} &
    \resizebox{88mm}{!}{\includegraphics{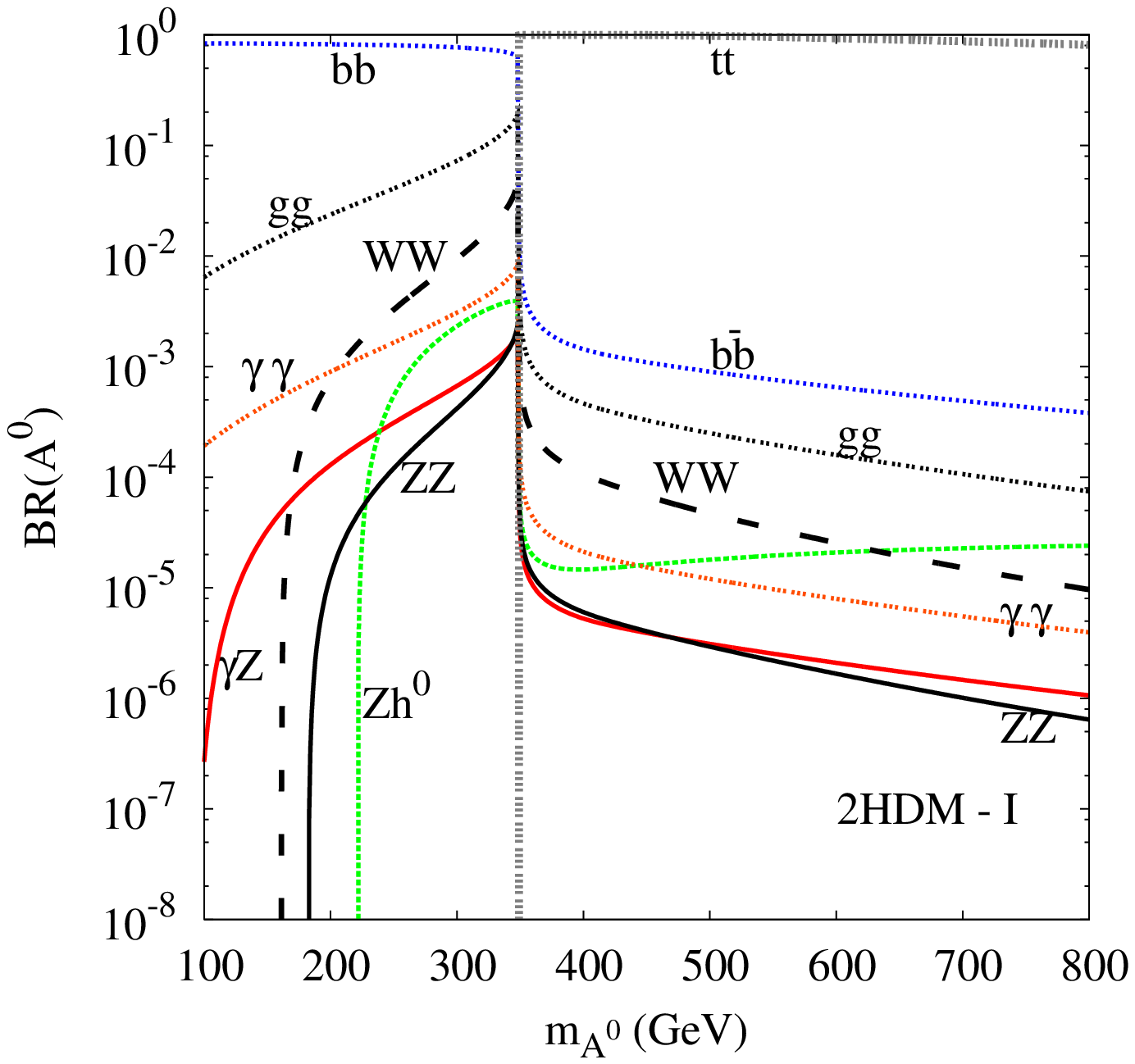}}
  \end{tabular}
\caption{Decays widths (left) and branching ratios (right) 
         of \textsc{cp}-odd $A^0$ as a
         function of  $m_{A^0}$ in 2\textsc{hdm} model type I,
         with parameters same as in Fig.\ref{fig:fig1}.}
\label{fig:fig2}
\end{figure}
\begin{figure}
  \begin{tabular}{cc}
    \resizebox{88mm}{!}{\includegraphics{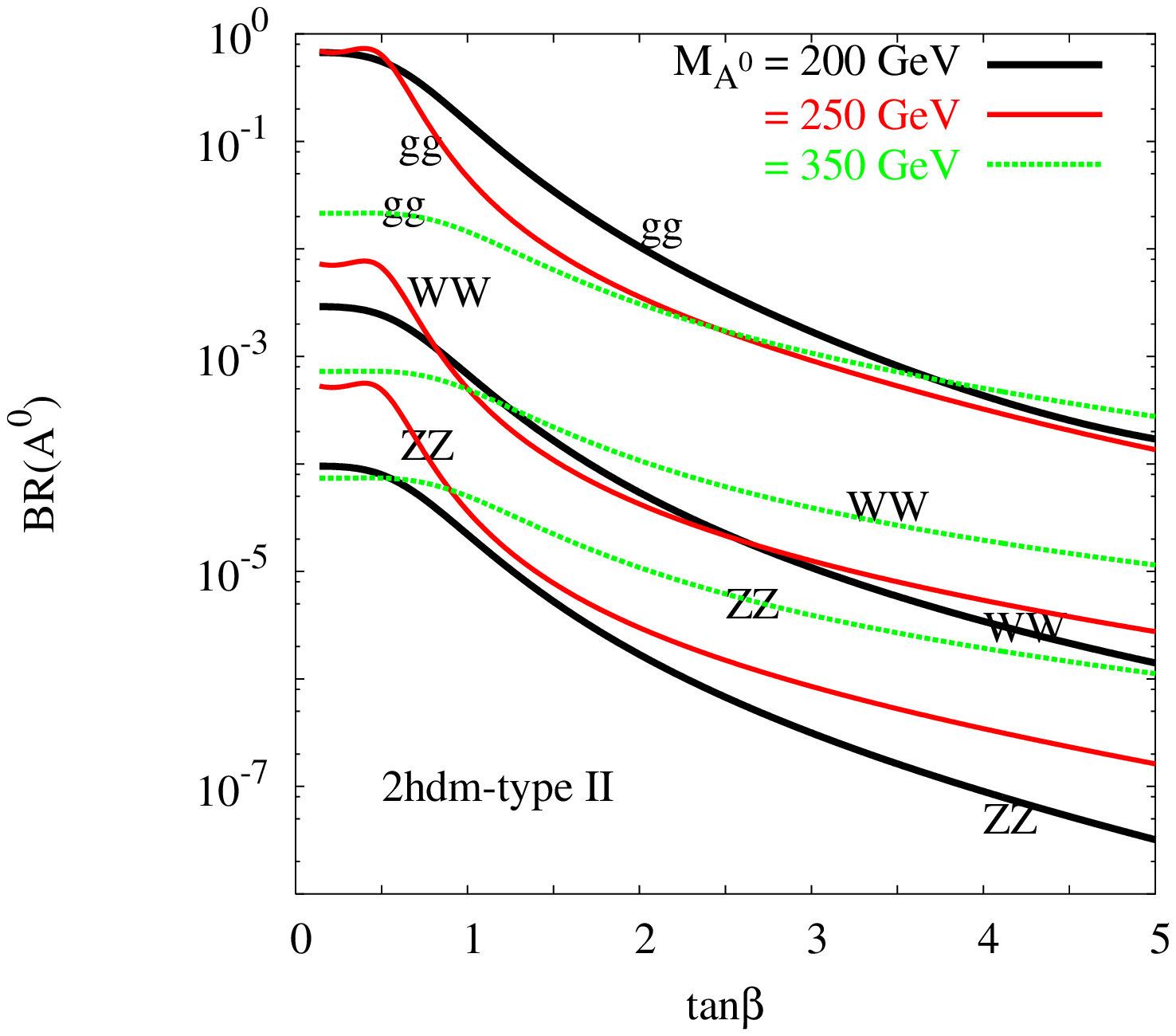}} &
    \resizebox{88mm}{!}{\includegraphics{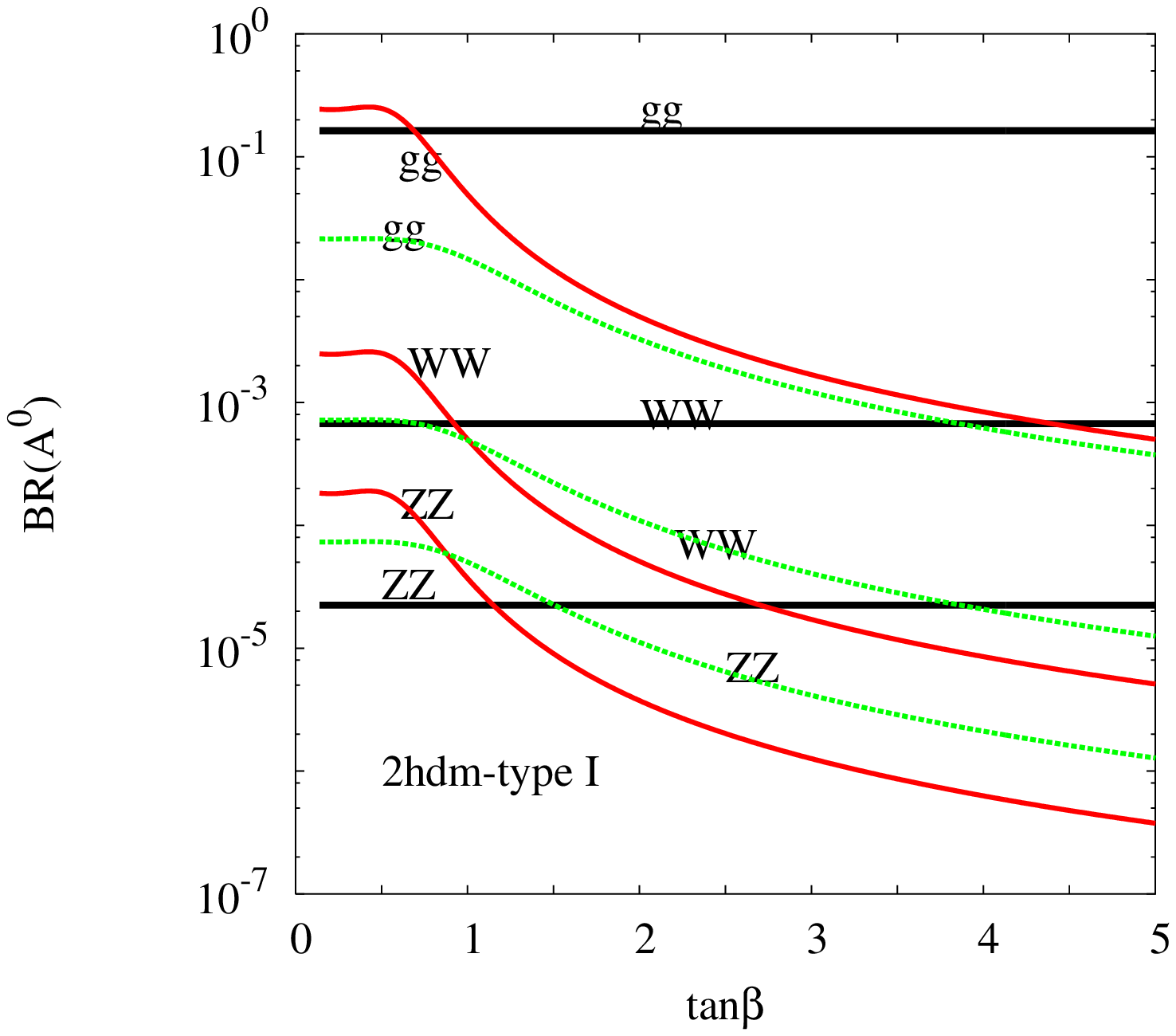}}
  \end{tabular}
\caption{Branching ratios of \textsc{cp}-odd $A^0$ into
$ZZ$, $W^+ W^-$ and $gg$ as a function of $\tan\beta$ in
type II 2\textsc{hdm} model (left), type I 2\textsc{hdm}
model (right), for the parameters $m_{h^0} = 130$ GeV,
$m_{H^0}=345$ GeV, $m_{H^{\pm}}=340$ GeV,
 and for various values of  $m_{A^0}$.}
\label{fig:fig3}
\end{figure}
For illustration we show first in Fig.~\ref{fig:fig1} and Fig.~\ref{fig:fig2}
different partials decays widths of $A^0$ both in 2\textsc{hdm} 
type II and type I as a function of $m_{A^0}$ and for small $\tan\beta =0.42$.

In Fig.~\ref{fig:fig1} (left) we illustrate branching ratios
of $A^0$ as a function of $m_{A^0}$ for small $\tan\beta =0.42$ scenario in the
framework of  2\textsc{hdm} type II.
As it can be seen from the plot, for $m_{A^0} \la 200$ GeV, $b\bar{b}$ 
mode is the dominant decay mode, but when approaching $t\bar{t}$ threshold
the $A^0 \to gg$ mode is enhanced and becomes the dominant decay mode
for $m_{A^0}\in [200,350]$ GeV. In the case when $m_{A^0}\approx 2 m_t$,
one can say that the \textsc{cp}-odd is almost fermiophobic 
\cite{Gunion:1989we}. After crossing $t\bar{t}$ threshold,
$A^0 \to t\bar{t}$ becomes the dominant decay mode. In this scenario the
decays of our concerns $A^0 \to W^+W^-$ and $A^0 \to ZZ$ can be respectively 
of the order $10^{-1}$ and $8\times 10^{-3}$ near $t\bar{t}$ threshold
region and decreases away from this region.

In type I 2\textsc{hdm}, the situation is slightly different.
For small $\tan\beta\approx 0.42$, the coupling of $A^0$ to a pair of bottom
quark, because it is proportional to $\cot\beta$, is enhanced.
In this case, before $t\bar{t}$ threshold $m_{A^0}\la 2 m_t$, 
$b\bar{b}$  is indisputably the
dominant decay mode followed by $A^0\to gg$ mode which is in the range
$\approx 10^{-2}$-$10^{-1}$. Again, after crossing $t\bar{t}$ threshold,
$A^0 \to t\bar{t}$ becomes the dominant decay mode.
As one can see, near $t\bar{t}$ threshold the branching ratio 
of $A^0 \to W^+W^-$ and $A^0 \to ZZ$ are of the order 
$10^{-2}$ and $10^{-3}$ respectively.

In Fig.~\ref{fig:fig3} (left) we show branching ratios of $A^0$ 
as a function of $\tan\beta$ for various choice of $m_{A^0}$ both in 
2\textsc{hdm} type II (left) and 2\textsc{hdm} type I (right).
It is clear that in 2\textsc{hdm} type II and for small 
$\tan\beta \la 1$, $A^0\to gg$ is the dominant decay mode
for $m_{A^0} \la 2 m_t$. One could say that in this region the
\textsc{cp}-odd is almost fermiophobic. One can see that 
all the decays $A^0 \to VV$, $V=W,Z$ are enhanced for the small
$\tan\beta$ limit. This is mainly due to top Yukawa coupling
which is proportional to $\cot\beta$. For large $\tan\beta$ 
all the branching ratio $A^0 \to VV$ decreases. \\
We note finally that in 2\textsc{hdm} type I, the $A^0 \to gg$ decay mode
is independent of $\tan\beta$ for $m_{A^0}=200$ GeV. This is because for this 
value of $m_{A^0}=200$ GeV, only fermionic decay are open and then $\tan\beta$ 
Dependance drop in the ratio.

\begin{figure}[H]
  \begin{tabular}{cc}
    \resizebox{88mm}{!}{\includegraphics{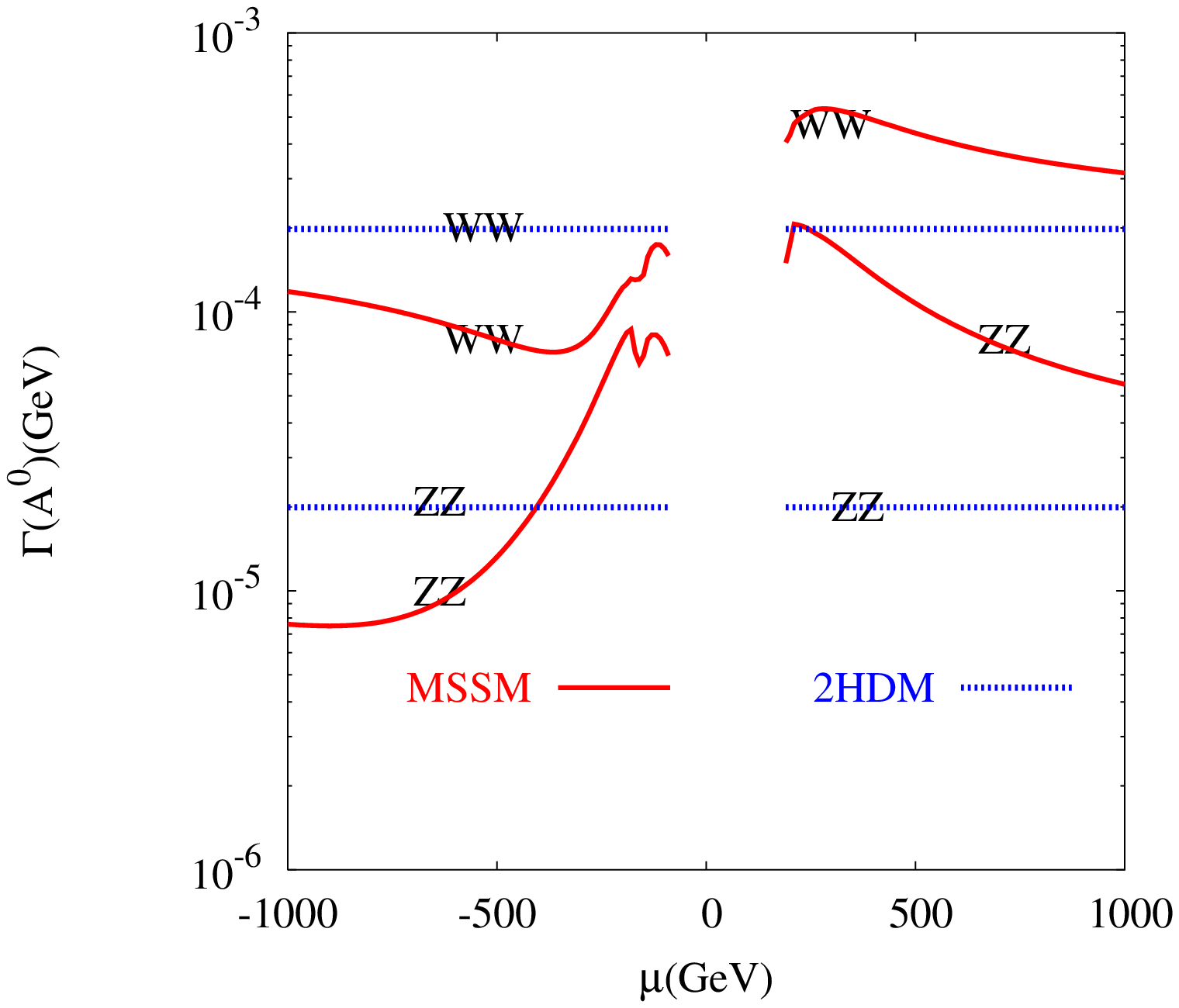}} &
    \resizebox{88mm}{!}{\includegraphics{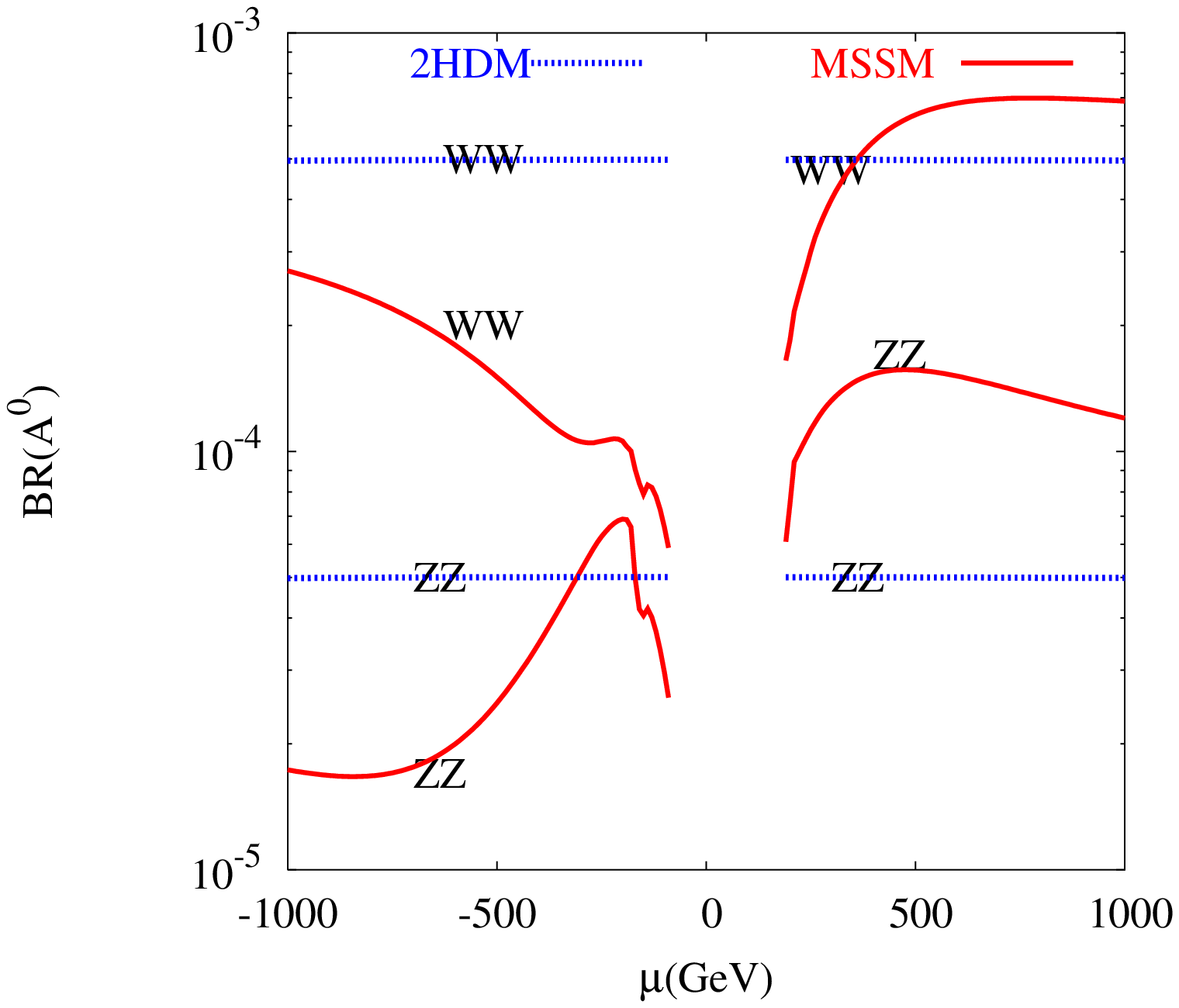}}
  \end{tabular}

\caption{Charginos and neutralinos contributions to the decays widths and
         branching ratios of $A^0\to ZZ$, and $A^0\to W^+W^-$ 
as function of $\mu$ parameter in
         \textsc{mssm} and Type-II 2\textsc{hdm} models, for the
         parameters $M_{Susy} = 500$ GeV, $M_2 = 140$ GeV, $m_{A^0} =
         350$ GeV, $\tan\beta = 5$, $A_t = 1$ TeV.}
\label{fig:fig4}
\end{figure}
In the framework of MSSM, we show first in Fig.\ref{fig:fig4}
the partials decays widths Fig.\ref{fig:fig4} (left) and branching ratios
 Fig.\ref{fig:fig4} (right) of $A^0$ into $W^+W^-$ and ZZ 
as a function of $\mu$ and compare it to its type II 2\textsc{hdm}. 
As it can be seen from this plot,
that chargino-neutralino contribution can either enhance the width 
(for Positif $\mu >0$) or suppress (for negative $\mu$).
It is clear from the right plot that the branching ratio of 
$A^0 \to W^+W^-$ (resp $A^0 \to ZZ$) can be of the order $10^{-3}$ 
(resp $10^{-4}$).

\begin{figure}[H]
  \begin{tabular}{cc}
    \resizebox{88mm}{!}{\includegraphics{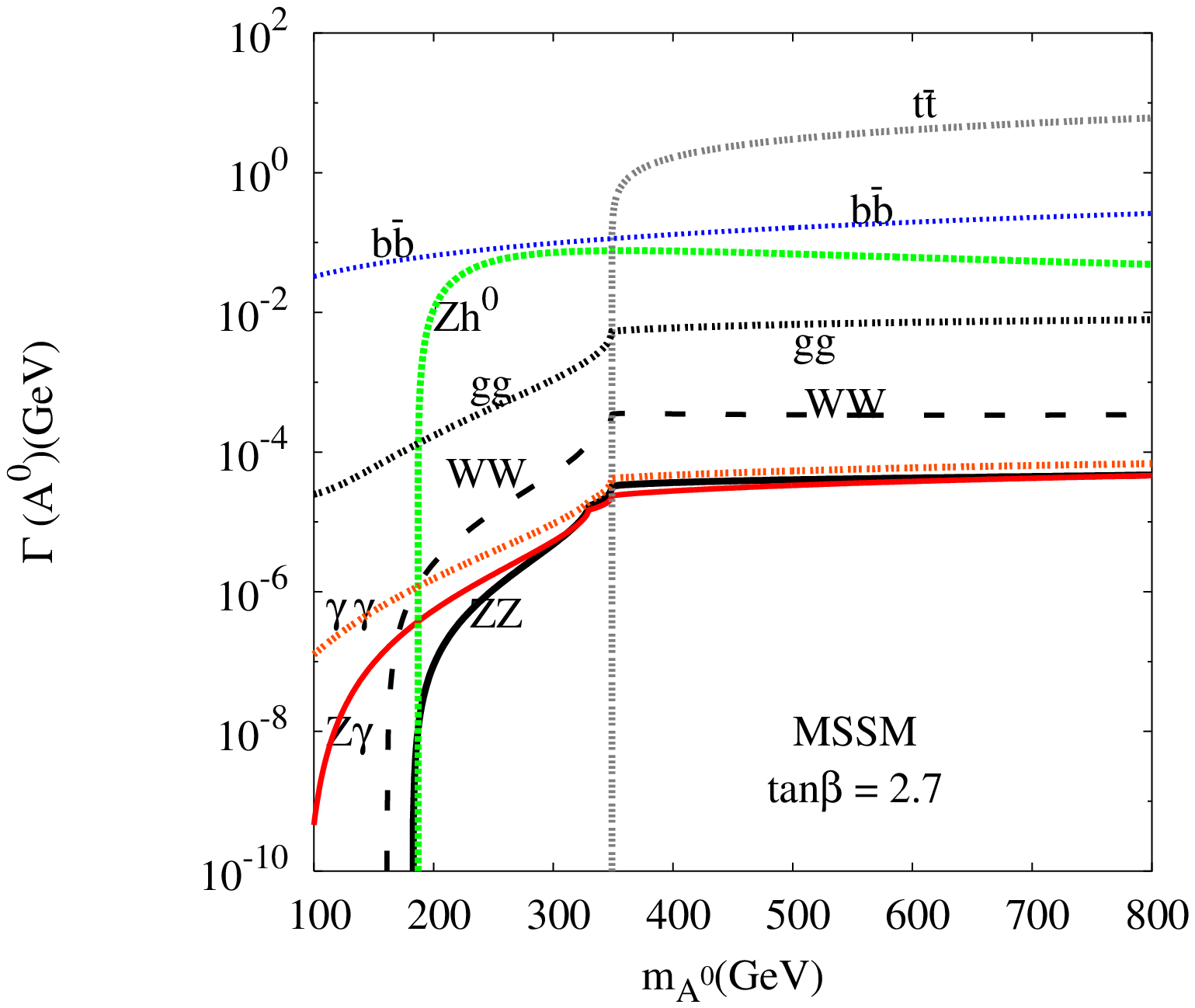}} &
    \resizebox{88mm}{!}{\includegraphics{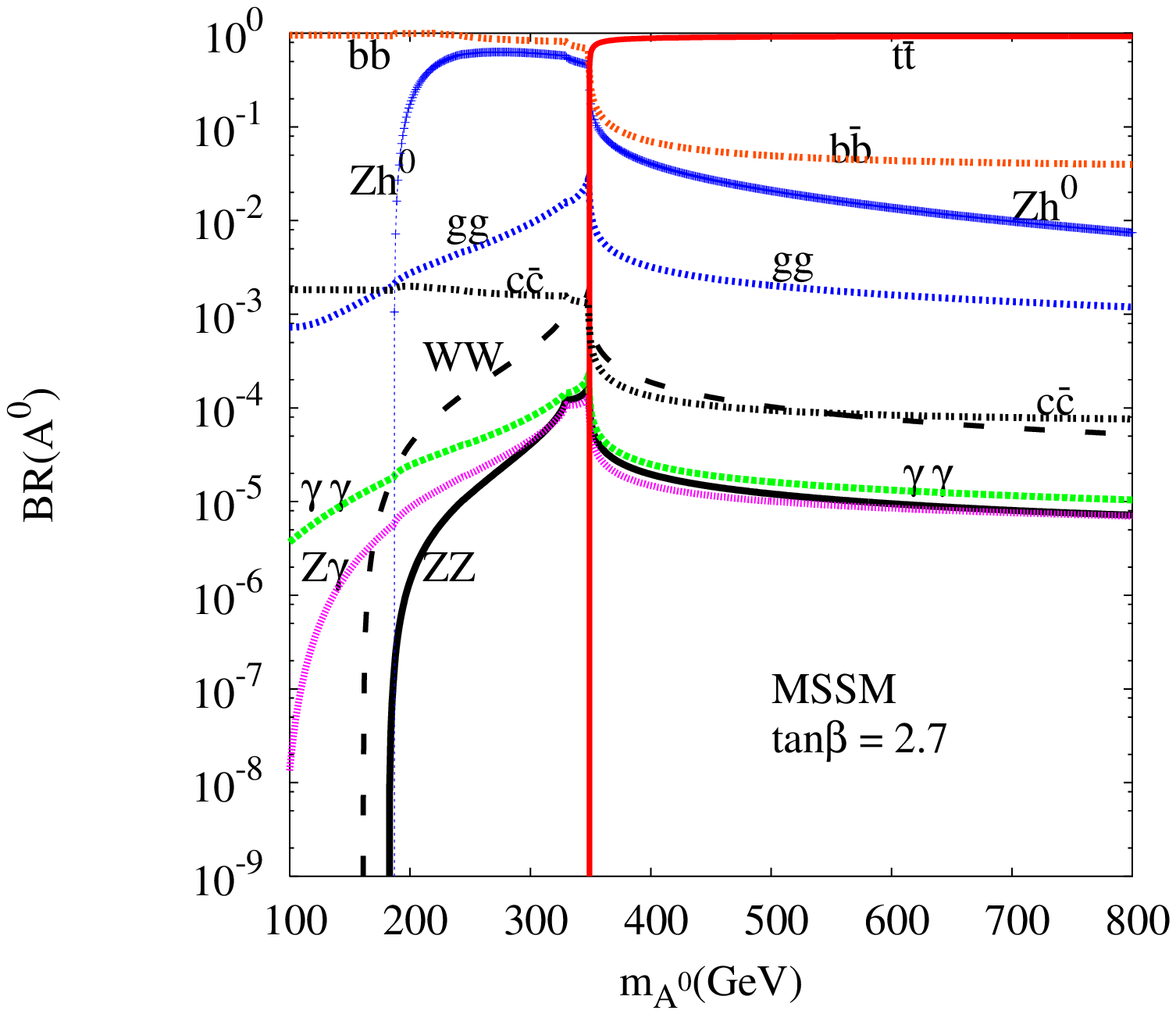}}
  \end{tabular}
\caption{Decays widths and branching ratios of 
         \textsc{cp}-odd $A^0$ in the \textsc{mssm} model as a function
    of $m_{A^0}$, for the parameters $M_{SUSY} = 500$ GeV, $M_2 =
         170$ GeV, $\mu = -A_t = 1$ TeV, $\tan\beta = 2.7$.}

\label{fig:fig5}
\end{figure}

\begin{figure}[H]
  \begin{tabular}{cc}
    \resizebox{88mm}{!}{\includegraphics{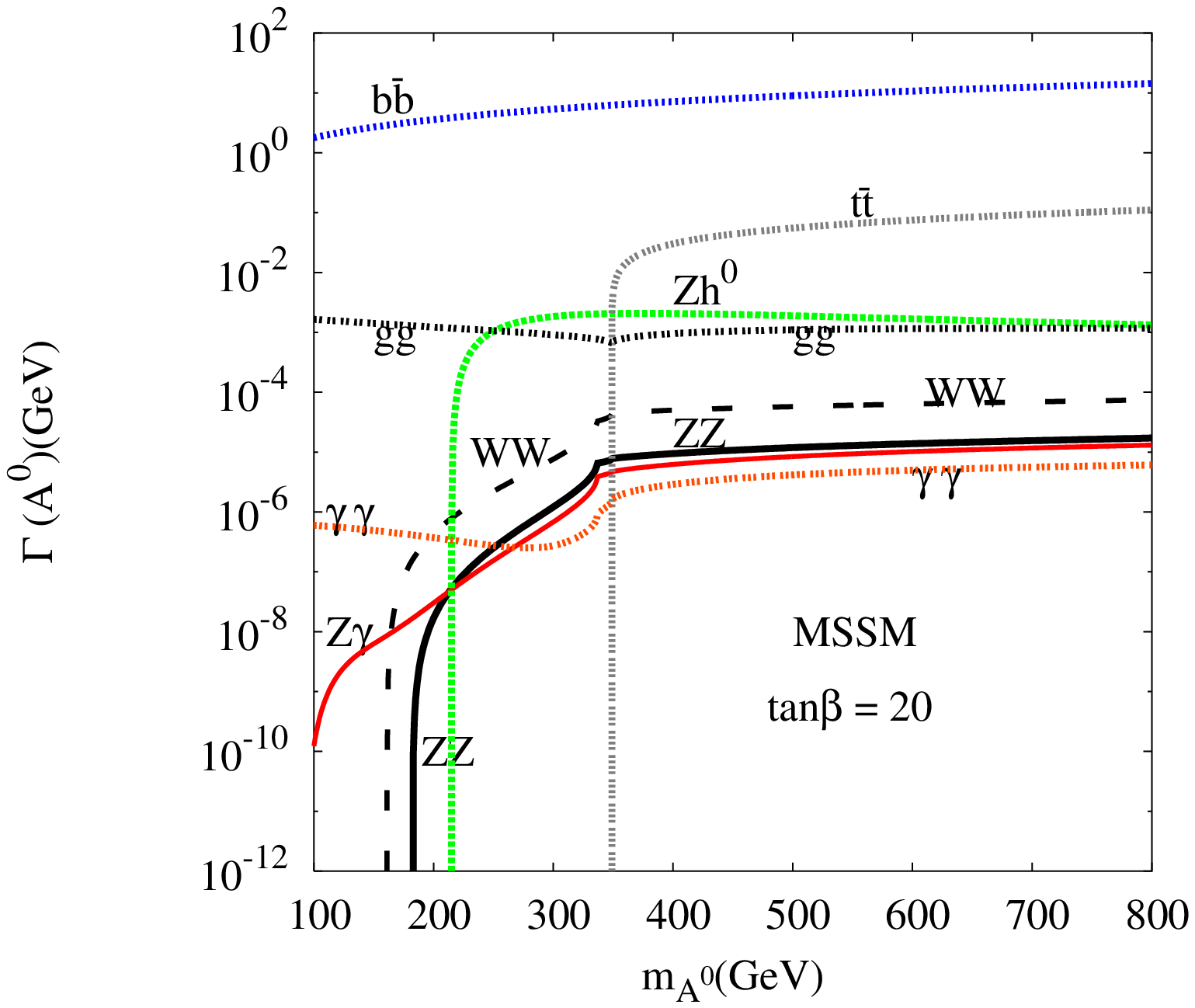}} &
    \resizebox{88mm}{!}{\includegraphics{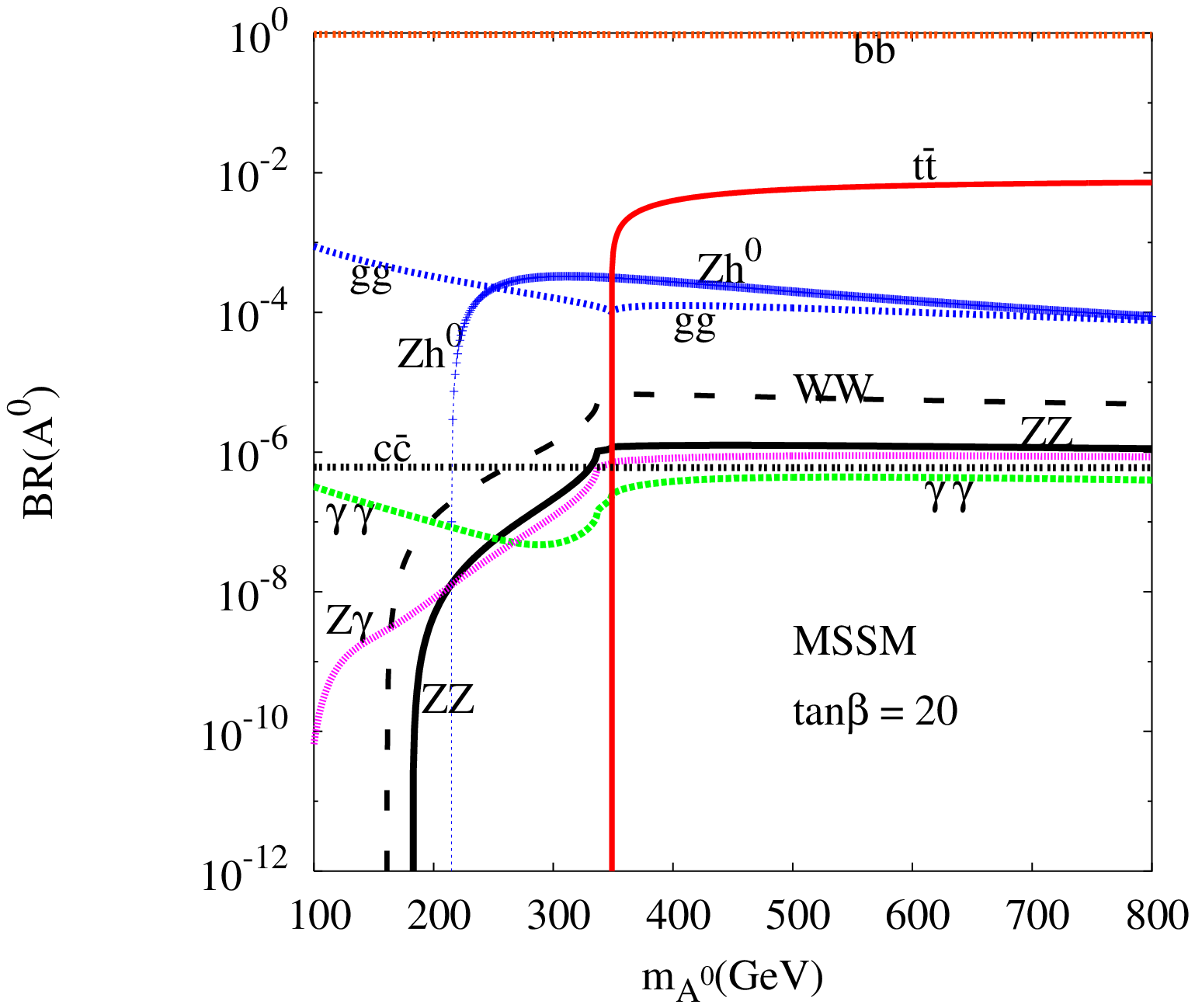}}
  \end{tabular}
\caption{Widths decays and branching ratios of Higgs boson
         \textsc{cp}-odd $A^0$ in the \textsc{mssm} model as function
         $m_{A^0}$, for the parameters $M_{SUSY} = 500$~GeV, $M_2 =
         170$~GeV, $\mu = -A_t = 1$~TeV, $\tan\beta = 20$.}

\label{fig:fig6}
\end{figure}

In Fig.~\ref{fig:fig5} and Fig.~\ref{fig:fig6} we fix $\mu$ to 1 TeV
and $M_2=170$ GeV and plot the decay width and branching ratio of $A^0$ 
as a function of $m_{A^0}$ for low $\tan\beta=2.7$ 
(Fig.~\ref{fig:fig5}) and $\tan\beta=20$  (Fig.~\ref{fig:fig6}).
For low $\tan\beta=2.7$, before the opening of $t\bar{t}$ mode for
$m_{A^0}\approx 350$ GeV, the dominant decay mode for $A^0$ is 
$b\bar{b}$ mode followed by $Zh^0$ mode for $m_{A^0}\ge 200$ GeV.
Once $t\bar{t}$ mode is open it is fully dominating. 
The branching ratio of $A^0\to WW$ (resp $A^0\to ZZ$) 
can reach a value of the order $10^{-3}$ (resp $10^{-4}$)
for \textsc{cp}-odd mass close to 2 $m_t$. However for large $\tan\beta=20$,
the situation is different, $b\bar{b}$ mode is fully dominating 
for all $m_{A^0}$ range. $A^{0}\to gg, Zh^0$ are at the level of $10^{-4}$ while 
$A^0\to W^+W^-$ and $A^{0}\to ZZ$ are at the level of $10^{-5}$ and $10^{-6}$
branching ratio respectively.
\section{Conclusions}
\label{sec:conclusion}
We have calculated the one-loop contribution to the $A^0WW$ and
$A^0ZZ$ couplings (absent at tree level) both in the 
context of Minimal Supersymmetric Standard Model \textsc{mssm} 
and Two Higgs Doublet Model 2\textsc{hdm}. 
In the \textsc{2hdm}, for small $\tan\beta$, 
near $t\bar{t}$ threshold, the branching ratio of $A^0 \to W^+W^-$ and, 
$A^0 \to ZZ$ are enhanced to the level of $10^{-1}$ and $10^{-2}$
respectively. We stress also that in 2\textsc{hdm} type II and
near $t\bar{t}$ threshold region,
the branching ratio of $A^0 \to gg$ could dominate aver $A^0\to b\bar{b}$ mode
and hence $A^0$ becomes fermiophobic.\\
In the MSSM, the branching ratios of 
$A^0 \to W^+W^-$ and $A^0\to ZZ$ are of the order
$10^{-3}$ and $10^{-4}$ respectively  for low $\tan\beta$
and decrease for large $\tan\beta$.
Those branching ratios, at this level,  might provide an opportunity to
search for a {\sc{cp}}-odd Higgs boson at the {\sc{lhc}}.

\begin{acknowledgments}
This work is supported in part by the U.S. Department of Energy under
grant DE-FG02-97ER41022. This work is supported by PROTARS-III D16/04.

\end{acknowledgments}
\appendix
\section{One-loop amplitudes for $A^0\to VV$}
\label{app:one-loop}

Let us briefly recall the definitions of scalar and tensor integrals
we use. We use the convention of LoopTools\cite{vanOldenborgh:1990yc,
Hahn:1999mt}.  The inverse of the propagator are denoted by,
\begin{equation} 
d_0 = q^2 - m_0^2 \ , \ d_i= (q+p_i)^2-m_i^2
\end{equation}
where the $p_i$ are the momenta of the external particles (always
incoming).

\begin{equation} 
B_{0,\mu} = \frac{(2\pi \mu)^{(4-D)} }{i \pi^2}
\int d^Dq \frac{\{ 1,q_{\mu} \}}{d_0 d_1} 
\end{equation}
Using Lorentz covariance, one gets for the vector integral 
\begin{equation} 
B_\mu  =  {p_1}_{\mu} B_1
\end{equation}
with the scalar function $B_1(p_1^2, m_0^2, m_1^2)$.

For the three-point function we have,
\begin{equation}
 C_{0,\mu}  = 
\frac{(2\pi \mu)^{(4-D)} }{i \pi^2}
\int d^Dq \frac{ \{1, q_{\mu},q_{\mu}q_{\nu}\} }{d_0 d_1 d_2}
\end{equation}
where $p_{12}^2=(p_1 + p_2)^2$. Lorentz covariance yields the
decomposition
\begin{equation}
C_\mu  =  {p_{1\mu} } C_1 +  p_{2\mu} C_2 
\end{equation}
with the  scalar functions 
$C_{i}(p_1^2,p_{12}^2,p_2^2,m_0^2,m_1^2,m_2^2)$.

\section{Lagrangian and Couplings}
\label{app:lagrangian}

In this Appendix, we use the fermion-vector boson coupling constants
as defined in terms of the neutral-current and charged-current
vertices.
\begin{eqnarray} \nonumber
{\cal L} &=& g \bf{W}^{-}_{\mu}
\left\{
   \widetilde{\chi}_j^{-} \gamma^{\mu} 
   \Big( \cO^{L}_{ji} P_{L} + \cO_{ji}^{R} P_{R} \Big)
   \widetilde{\chi}_i^{0}
   + \frac{1}{\sqrt{2}}
   \Big( \bar{l}_{L_i} \gamma^{\mu} P_{L} \nu_{L_i}
       + V^{ij}_{CKM} \bar{d}_{L_j} \gamma^{\mu} P_{L} u_{L_i} 
   \Big) + h.c 
\right\} \\ \nonumber
         &+& g\sin\theta_W  \bf{A}_{\mu}
\Big( \widetilde{\chi}_i^{+} \gamma^{\mu} 
      \widetilde{\chi}_j^{-} + h.c
\Big) \\ \nonumber
         &+& \frac{g}{\cos\theta_W} \bf{Z}_{\mu}
\Big[ \frac{1}{2}\widetilde{\chi_i}^{0} \gamma^{\mu}
   \Big( {\cO''}^{L}_{ji} P_{L} 
       + {\cO''}_{ji}^{R} P_{R} 
   \Big)
   \widetilde{\chi}_j^{0} + \widetilde{\chi}_i^{+} \gamma^{\mu}
   \Big({\cO'}^{L}_{ij} P_{L} + {\cO'}_{ij}^{R} P_{R} 
\Big) \widetilde{\chi_i}^{-} \\
         &+& \sum_{f=l,u,d} \bar{f}
\Big( (-I_{f}^3 + Q_{f}\sin^{2}\theta_W  )P_L 
                + Q_{f}\sin^{2}\theta_W P_R \Big) 
\Big]
\end{eqnarray}
where the real $4\times 2$-matrices ${\cO}^{L,R}_{ij}$, symmetric and
real $2\times 2$-matrices ${\cO'}^{L,R}_{ij}$, and symmetric and real
$4\times 4$-matrices ${\cO''}^{L,R}_{ij}$ have the form:
\begin{eqnarray}
\cO_{ji}^{L}    &=& N_{i2}U_{j1}-\frac{1}{\sqrt{2}}N_{i3}U_{j2} \\
\cO_{ji}^{R}    &=& N_{i2}V_{j1}+\frac{1}{\sqrt{2}}N_{i4}V_{j2} \\
{\cO'}_{ij}^{L} &=& -\frac{1}{2} U_{i2} U_{j2}
                    - U_{i1} U_{j1} + \sin^2\theta_W\delta_{ij} \\
{\cO'}_{ij}^{R} &=& -\frac{1}{2} V_{j2} V_{i2}
                    - V_{j1} V_{i1} + \sin^2\theta_W\delta_{ij} \\
{\cal O''}_{ij}^{L} &=&  \frac{1}{2}\Big( N_{i4} N_{j4} 
                                        - N_{i3} N_{j3} \Big)   \\
{\cal O''}_{ij}^{R} &=& -\Big( {\cal O''}_{ij}^{L} \Big)
\end{eqnarray}
The Lagrangian Higgs \textsc{cp}-odd neutralinos and charginos
\begin{equation}
{\cal L} = -ig 
\left\{
    \sum_{i,j=1}^{4} \bar{\tilde{\chi}}^0_{i}
    \Big( {\cal O}^{NNA}_{ij} P_R + {\cal S}^{NNA}_{ij} P_L
    \Big) \tilde\chi^0_{j} A^{0}
  + \sum_{i,j=1}^{2} \bar{\tilde{\chi}}^{\pm}_{i}
    \Big( {\cal O}^{CCA}_{ij} P_L + {\cal S}^{CCA}_{ij} P_R
    \Big) \tilde\chi^{\pm}_j A^0
\right\}
\end{equation}
where
\begin{eqnarray}
{\cal O}^{NNA}_{ij} &=& \frac{1}{2}
        \left\{
             \Big( \sin\beta N_{i3} - \cos\beta N_{i4} \Big)
             \Big( \sin\theta_W N_{j1} - \cos\theta_W N_{j2} \Big) +
             (i \leftrightarrow j)
        \right\} \\
{\cal S}^{NNA}_{ij} &=& \Big( {\cal O}^{NNA}_{ji} \Big) \\
{\cal O}^{CCA}_{ij} &=& \frac{1}{\sqrt{2}}
        \Big( \sin\beta U_{i2} V_{j1} + \cos\beta U_{i1} V_{j2} \Big)\\
{\cal S}^{CCA}_{ij} &=& -\Big( {\cal O}^{CCA}_{ji} \Big)
\end{eqnarray}

\section{One-loop amplitude}
\label{app:amplitude}
In \textsc{mssm} and 2\textsc{hdm}-II, the amplitude of the sum of
diagrams Fig.~\ref{fig:azz-diagrams}.a and \ref{fig:azz-diagrams}.b,
is given by:
\begin{equation}
{\cal A}_{VV} = {\cal A}_{VV}^{2\textsc{hdm}} + {\cal A}_{VV}^{\textsc{susy}}
\end{equation}
\begin{itemize}
\item $A^{0}\to ZZ$
\end{itemize}
\begin{eqnarray}
{\cal A}_{ZZ}^{2\textsc{hdm}} &=& i \frac{2}{c^2_W}
   \sum_{f=u,d} m^{2}_{f} g_{A^0\bar f f}
   \left\{ \Big( (g^{R}_{Z\bar ff})^2 + (g^{L}_{Z\bar ff})^2 \Big)
           C_{0}^{f} 
         + \Big( g^{R}_{Z\bar{f}f} - g^{L}_{Z\bar{f}f} \Big)^2 C_{1}^{f}
   \right\} \\
{\cal A}_{ZZ}^{2\textsc{susy}} &=& i \frac{4m_W}{c^3_W}
   \sum_{i,j,k=1}^{4} {\cO''}^{L}_{ik}
                      {\cO''}^{L}_{jk}
                      {\cO}^{NNA}_{ij}
   \left\{  m_{\tilde{\chi}^{0}_i}   C^{ikj}_0 
        + ( m_{\tilde{\chi}^{0}_i}
        +   m_{\tilde{\chi}^{0}_k} ) C^{ikj}_1 
        + ( m_{\tilde{\chi}^{0}_i}
        -  m_{\tilde{\chi}^{0}_j}  ) C^{ikj}_2
   \right\} \\ \nonumber
                     &-& i \frac{4m_W}{c^3_W}
   \sum_{i,j,k=1}^{2}
   \Big\{ m_{\tilde{\chi}^{\pm}_i} ( {\cal O'}_{jk}^{R}
                                     {\cal O'}_{ik}^{R}
                                     {\cal O}_{ij}^{CCA}
                                   - {\cal O'}_{jk}^{L}
                                     {\cal O'}_{ik}^{L}
                                     {\cal S}_{ij}^{CCA} )
                     ( C_0^{ikj} + C_1^{ikj} + C_2^{ikj} ) \\
   \nonumber 
                     &-& m_{\tilde{\chi}^{\pm}_k}( {\cal O'}_{jk}^{R}
                                                   {\cal O'}_{ik}^{L}
                                                   {\cal O}_{ij}^{CCA}
                                                 - {\cal O'}_{jk}^{L} 
                                                   {\cal O'}_{ik}^{R}
                                                   {\cal S}_{ij}^{CCA} )
                                               C_1^{ikj} \\ \nonumber
                     &-& m_{\tilde{\chi}^{\pm}_j}( {\cal O'}_{jk}^{L}
                                                   {\cal O'}_{ik}^{L}
                                                   {\cal O}_{ij}^{CCA}
                                                 - {\cal O'}_{jk}^{R}
                                                   {\cal O'}_{ik}^{R}
                                                   {\cal S}_{ij}^{CCA} )
                                               C_2^{ikj} \Big\}
\end{eqnarray}
Where $g_{A^0\bar f f} = \tan\beta (\cot\beta)$, for up fermion (down
fermion). The arguments of $C_{0,1}^f$ are $(m_Z^2, m_Z^2, m_{A^0}^2,
m_f^2, m_f^2, m_f^2)$ and $C_{0,1,2}^{ikj}$ are $(m_Z^2, m_Z^2,
m_{A^0}^2, m_{\tilde{\Psi}_i},m_{\tilde{\Psi}_k},m_{\tilde{\Psi}_j})$,
where $\Psi_a = \tilde\chi^{0}_i$ or $ \tilde\chi^{\pm}_i$.

\begin{itemize}
\item $A^{0} \to W^+W^-$
\end{itemize}
\begin{eqnarray}
{\cal A}_{WW}^{2\textsc{hdm}} &=& i \sum_{f,f'=u,d}
    m^2_f g^{2}_{W\bar{f}' f} g_{A^{0}\bar{f}f}
    \Big( C_{0}^{ff'f} + C_{1}^{ff'f} \Big) \\
{\cal A}_{WW}^{2\textsc{susy}} &=& i \frac{2m_W}{c_W}
    \sum_{i,j,k=1}^{4} {\cO}^{NNA}_{ij}
    \bigg\{ \Big( {\cO}^{R}_{ki} {\cO}^{R}_{kj}
                + {\cO}^{L}_{ki} {\cO}^{L}_{kj} \Big)
    \bigg[ m_{\tilde{\chi}^{0}_i} ( C_0 + C_1 ) 
        + (m_{\tilde{\chi}^{0}_i} - m_{\tilde{\chi}^{0}_j} ) C_2
    \bigg] \\
                    &-&
           m_{\tilde{\chi}^{\pm}_k}
    \Big( {\cO}^{R}_{kj} {\cO}^{L}_{ki} + {\cO}^{L}_{kj} {\cO}^{R}_{ki}
    \Big) C_1
    \bigg\}\\ \nonumber
                    &-& i 2 m_{W}
    \sum_{\substack{i,j=1,2 \\ k=1,4}}
    \bigg\{ m_{\tilde{\chi}^{\pm}_i} 
      \big( {\cO}^{R}_{jk} {\cO}^{R}_{ik} {\cO}^{CCA}_{ij}
          - {\cO}^{L}_{jk} {\cO}^{L}_{ik} {\cS}^{CCA}_{ij} \big)
      \big( C_0 + C_1 + C_2 \big) \\ \nonumber
                    &+& 
            m_{\tilde{\chi}^{\pm}_j} 
      \big( {\cO}^{R}_{jk} {\cO}^{R}_{ik} {\cS}^{CCA}_{ij}
          - {\cO}^{L}_{jk} {\cO}^{L}_{ik} {\cO}^{CCA}_{ij} \big) C_2
          + m_{\tilde{\chi}^{0}_k}
      \big( {\cO}^{L}_{jk} {\cO}^{R}_{ik} {\cS}^{CCA}_{ij}
          - {\cO}^{R}_{jk} {\cO}^{L}_{ik} {\cO}^{CCA}_{ij} \big) C_1
    \bigg\}
\end{eqnarray}
where the $C_{0,1}^{ff'f}$ have the same arguments
$C^{ff'f}_{0,1}(m_W^2$, $m_W^2$, $m_{A^{0}}$, $m^2_f$, $m^2_{f'}$,
$m^2_f)$, all the $C_{0,1,2}$ have also the arguments
$C_{0,1,2}(m_W^2$, $m_W^2$,
$m_{A^{0}},m_{\tilde{\Psi}},m_{\tilde{\Psi}},m_{\tilde{\Psi}})$ where
$\tilde{\Psi} = \tilde{\chi}^{\pm}_i$ or $\tilde{\chi}^{0}_i$.

\begin{itemize}
\item $A^{0} \to Z \gamma$
\end{itemize}
\begin{eqnarray}
{\cal A}_{Z\gamma}^{2\textsc{hdm}} &=& -i 2\tan\theta_W
   \sum_{f=d,u} N_C \, Q_{f} \, m^{2}_{f} g_{A^{0}\bar{f}f}
   \Big( g^{L}_{Z\bar{f}f} + g^{R}_{Z\bar{f}f} \Big) \, C_0 \\ \nonumber
{\cal A}_{Z\gamma}^{\textsc{susy}} &=& -i 2 m_{Z}\sin\theta_W
   \sum_{i,j=1}^{2} ( {\cal O'}_{ij}^{L} + {\cal O'}_{ij}^{R} )
                    ( {\cal S}_{ij}^{CCA} - {\cal O}_{ij}^{CCA} )
   \left\{ m_{\tilde{\chi}^{\pm}_i} \ ,C_0
        +( m_{\tilde{\chi}^{\pm}_i}
        - m_{\tilde{\chi}^{\pm}_j} ) \, C_2
   \right\}
\end{eqnarray}
where $C_{0,2}$ have the same arguments $C_{0,2}(0, m_Z^2, m_A^2,
m^2_{\tilde{\chi}^{\pm}_i}, m^2_{\tilde{\chi}^{\pm}_i},
m^2_{\tilde{\chi}^{\pm}_j})$.

\begin{itemize}
\item $A^{0} \to \gamma \gamma$
\end{itemize}
\begin{eqnarray}
{\cal A}_{\gamma\gamma}^{2\textsc{hdm}} &=& - i4
  \sum_{f=d,u} N_C \, Q^2_{f} \, m^{2}_{f}\sin^2\theta_W \, 
               g_{A^{0}\bar{f}f} C_{0} \\ \nonumber
{\cal A}_{\gamma\gamma}^{\textsc{susy}} &=& -i 8m_W \, \sin\theta_W
  \sum_{i=1,2} m_{\tilde{\chi}^{\pm}_i}{\cO}^{CCA}_{ii} C_{0}
\end{eqnarray}
The $C_0$ have the argument $C_{0}(0, 0, m_A^2,
m^2_{\tilde{\chi}^{\pm}_i}, m^2_{\tilde{\chi}^{\pm}_i},
m^2_{\tilde{\chi}^{\pm}_i})$.

\end{document}